# Self-organization principles of cell cycles and gene expressions in the development of cell populations

Xiaoliang Wang[1,2]*, Dongyun Bai[3]*

1 College of Life Sciences, Zhejiang University, Hangzhou 310058, China
2 School of Physical Sciences, University of Science and Technology of China, Hefei 230026, China
3 School of Physics and Astronomy, Shanghai Jiao Tong University, Shanghai 200240, China

*Correspondence: (X.W.) wxliang@mail.ustc.edu.cn; (D.B.) bdy1993@sjtu.edu.cn;

## Abstract

A big challenge in current biology is to understand the exact self-organization mechanism underlying complex multi-physics coupling developmental processes. With multiscale computations of from subcellular gene expressions to cell population dynamics that is based on first principles, we show that cell cycles can self-organize into periodic stripes in the development of *E. coli* populations from one single cell, relying on the moving graded nutrient concentration profile, which provides directing positional information for cells to keep their cycle phases in place. Resultantly, the statistical cell cycle distribution within the population is observed to collapse to a universal function and shows a scale invariance. Depending on the radial distribution mode of genetic oscillations in cell populations, a transition between gene patterns is achieved. When an inhibitor-inhibitor gene network is subsequently activated by a gene-oscillatory network, cell populations with zebra stripes can be established, with the positioning precision of cell-fate-specific domains influenced by cells' speed of free motions. Such information may provide important implications for understanding relevant dynamic processes of multicellular systems, such as biological development.

**Keywords:** cell-fate decision; multiscale computations; positional information; self-organization; zebra stripe

## 1. Introduction

The most fascinating thing in developmental biology is how a highly structured multicellular organism is developed from an initially homogenous zygote. Such a biological patterning is an intricate process that takes place at multiple time and space scales, involving cell division, cell differentiation and specification (fate determination), cell migration, cell interactions and even cell death [1]. To clearly understand such complex tissue-level patterning processes from behaviors of individual cells is a long-standing technical challenge for both experiments and theory.

    As early as 1952, Turing became the first to propose a mathematical model (i.e. the reaction-diffusion model) to explain morphogenesis [2]. Local self-enhancement and long-range inhibition that are based on the diffusing and reacting chemicals (usually termed morphogens), have been derived and widely believed as the fundamental mechanism for biological pattern formation [1-12] due to their ability to explain self-regulation and scaling of patterning, which are key characteristics of developmental biology [13]. With the increasingly deeper understanding of cellular activities, the Turing mechanism is being found far from adequate. A big problem of Turing models is that these models cannot directly treat individual behaviors of cells and have neglected the multi-scale aspect of biological pattern formation. This makes them hard to gain wide acceptance among experimental biologists, due to the gap with the real case [14]. Therefore, for current pattern formation theories, more precise and quantitative multiscale models are in great need and being

created [15-17].

In experiments, to have a clear understanding of biological pattern formation, some scientists utilized a bottom-up approach to model pattern formation in cell populations, by designing and building simple gene-regulatory circuits in some simple model microorganisms [18-22]. Such an approach has greatly reduced the difficulty in understanding biological patterning principles. However, there is still a long way for this method to go to disclose the mechanisms behind many more complex developmental processes, which also strongly needs the involvement of more powerful pattern formation models.

Here, we present first-principle computations to further explore the pattern formation principle in cell populations naturally growing on a two-dimensional plane. We select *E. coli.* as the model cell in our multiscale discrete element model (DEM, see Supporting Information for methods and the calibration with available experimental data [23]) incorporating intracellular gene expressions. In this study, we mainly stress two basic developmental questions: (1) How can a highly ordered spatial organization of cells be self-established within a cell population? (2) How can cell-fate-specific domains precisely positioning inside a cell population? The first question is explored in Sections 2.1-2.3, and the second is explored in Section 2.4. Our ultimate intention here is to provide possible new insights into relevant biological development through our explorations.

## 2. Results

This section orderly presents the patterning regimes of cell cycles, periodic gene expressions and zebra stripes. We will see later that some parameters used in our multiscale physical model are tunable in a range and are not based on available experimental values (see Table 1 for the parameters defined in this article). However, this is a valid approach as the focus of this article is to show the potential behaviors of models (the possible patterning ways) and not to study their predictive power using specific experimental values. Some critical predictions produced by our theoretical models remain to be revealed through future experiments.

### 2.1 Spatial organization of cell cycles

The spatial arrangement of cell fates within developing cell populations requires a clear understanding of population dynamic behaviors of individual cell cycles in the process. Growth and division cycle is the most basic behavior of cells, with a number of physiological activities strongly linked. Fig. 1a illustrates several phases within one cell cycle of one *E. coli* cell under the ideal steady-state growth condition, which precisely correspond to specific cell lengths (or cell mass) (Figs. 1b.c) [24] and physiological states, such as DNA replication levels [25,26]. Fig. 1c shows the correspondence between cell length $l_C$ and cell cycle phase $\varphi_C$ within multiple cell cycles. The correlation between them is given by:

$$\varphi_C = 2\pi \left( \frac{l_C - l_0}{l_{div} - l_0} \right), \quad l_0 \leq l_C \leq l_{div} \quad (1)$$

where $l_0$ is the initiation length of one cell cycle and $l_{div}$ is division length of cells (or the ending length of one cell cycle).

Although some previous works were performed on the interplay between single cell growth cycle and intracellular gene expressions [28-31], little is known about their spatial population dynamics and properties under the impact of their living environment. Actually, during the developing process from one single cell to tissue-level cell populations, growth resources (nutrients) that are available to individuals are often spatially heterogeneous due to their limited diffusion [25,32]. Growth resources have a direct determination of cell cycle time via influencing the synthesis rate of substances. Consequently, the spatial heterogeneity of resources will lead to the differentiation between cells' fates inside a population [25]. In this study, we first explore the effect of spatiotemporal self-allocation of growth resources on the population dynamics of cell cycles within expanding cell populations.

We grow one single cell on the solid agar under standard experimental conditions (see Fig. 2 for relevant parameters). We observe that when nutrients are unlimited, cell cycles cannot be coordinated into the spatially structured packing during the development of the cell population (Fig. 2a), since no graded position information is provided to cells (all cells have the identical division cycle time). In such a case, cell cycles are independent of each other. Because of the random effect naturally happening to cell divisions of all the cells, the spatial arrangement of cell cycle phases within the population is always disordered.

Under limited nutrients, cell cycles start to self-organize into periodic stripe structures after the colony grows to a specific size (Fig. 2b1, Movie S1), because in such a case, cells can obtain the coordinated position information provided by the spatiotemporal self-allocation of nutrients to lock into region-specific fates. During the early development stage (0~7.0 h), reduced nutrients available to inner cells can be supplemented timely through the diffusion of those at outside, so there is no big difference in cells' growth conditions (Fig. 2b3) and subsequent cells' growth rates (Fig. 2b2). With the further increase of colony size, however, inner cells are unable to obtain equally fair allocation of nutrients to keep pace with outer cells in cell division cycle phases due to the limited diffusion, which leads to the sufficiently graded radial position information and subsequent differentiation between inner and outer cells' cycle phases (t > 10h, Movies S1 to S3). Since the provided position information is radially symmetrical (Fig. 2b3), stripes are always the closed curves encircling the center point of the population.

We next explore the initiation condition of patterning. Fig. 3a shows that during the development of the cell population, self-organized patterning initiates at a sufficiently sloped radial nutrient distribution. We select the parameter $C_{front}/C_{center}$, the proportion of the nutrient concentration at the colony front to that at the center point of colony, to quantify this patterning condition. Fig. 3b shows that patterning starts once the proportion exceeds a specific critical value $k_c \approx 3$. Simulations under various growth conditions further indicate that patterning generally needs higher $k_c$ at higher nutrient concentrations and uptake rates and lower nutrient diffusion constants (Figs. 3c-e).

We further investigate the effects of some growth variables on the properties of stripe structures. Results show that an increased nutrient concentration and diffusion speed can lead to an increment in the wave length of cell size stripes $L_\lambda$ (Figs. 4a,b), while the increase of nutrient uptake rate could cause a reduction in wave length (Fig. 4c). The shared outcome of variations of these variables is the change in the moving speed of nutrient distribution $v_n$ (Fig. 4d) and also the expansion speed of cell populations (Fig. S1). Simulations show there is a positive linear correlation between $L_\lambda$ and $v_n$. Phase diagrams (Figs. S2 and S3) also show that stripes of cell cycles can be established more readily under lower nutrient concentrations, higher nutrient uptake rates and slower diffusion of nutrients.

## 2.2 Statistical distribution of cell cycles

A simple characterization of the packing structure of cell states within the population is the statistical probability distribution of cell cycle phases. In our study, we observe that under unlimited nutrients, the terminal probability density distribution of cell cycle phases within the population is periodically time-variant, despite various cell cycle times (Figs. 5a-c, Movie S4). But under limited nutrients, the statistical distributions will gradually collapse to some specific curves, with the establishment of more stripes (Figs. 5d-f, Movie S5). Surprisingly, these curves have a universal distribution, which is independent of the wave length of stripe structures (i.e. cell cycle distributions have a scale-invariant dependence) and can be well described with a four-parameter empirical function (Figs. 5g,h):

$$P(f) = a(1 - be^{-\alpha f})e^{-\beta f} \quad (2)$$

where $f = \varphi_C / \langle\varphi_C\rangle$ is the normalized cell cycle phase. $\langle.\rangle$ indicates the average over all cells in the population. Optimal fitting values are obtained: $a = 0.7472$, $b = 0.6120$, $\alpha = 14.81$ and $\beta = 0.5704$. The function (2) captures the exponential tail at large $f$ and the peak of the distribution near $f = 0.2$.

Note that the scale-invariant function (2) can be derived after only several cell cycle stripes are established within the cell population (Figs. 5d-f). This may indicate that the statistical distribution within a colony will step into a steady state soon after the starting of the cell-cycle self-organization, although the colony expansion has not entered into a steady state (Fig. S1b, where under the limited nutrients the colony expansion speed continues to decline with time to zero). Besides, the peak of the universal distribution locates at $f = 0.2 \ll 1$, which means within one complete cell cycle stripe, a high proportion of cells are fixed at the initial cell cycle phase. Moreover, the collapse property of the statistical distribution of cell cycles (e.g. the critical collapse speed) is likely to be used to assess if a real cellular system is starting to diverge away from the over-abundance of nutrients and enter into a cell-cycle self-organized regime.

## 2.3 Genetic pattern formation

Here we further explore the population dynamics of intracellular genetic oscillators, which may play a significant role in controlling various biorhythms and regulating biological pattern formation like color patterns on animal skins. Several types of synthetic gene oscillators have been built in model microorganisms [33,34] and the repressilator was recently shown in spatial patterning [21]. However, the exact patterning conditions and properties of gene oscillators still largely remain poorly understood due to the limitation of experimental techniques and the lack of proper pattern formation theories as well [13,35]. Here we use the multiscale DEM model to reveal underlying mechanisms of gene patterns in the development of cell populations, through incorporating intracellular gene expressions (Supporting Information).

We grow one single cell with the activator-inhibitor (AI) gene regulatory network which has been engineered in *E. coli* strains (Fig. 6a) on the solid agar (see Fig. 7 for parameters used for the current genetic kinetics within each cell). With this gene-oscillatory network, each cell is able to switch its states between the low and high gene expressions (Fig. 6b). Gene expressions are often strongly coupled with cell's growth state, to incorporate this effect, intracellular gene oscillators are therefore forced to be shut down when cells' growth rate reduces to a specific value $\lambda_{stop}$. In our simulations, $\lambda_{stop} = 0.5$ h$^{-1}$ unless otherwise stated.

We observed three modes of genetic patterns under various growth conditions (different nutrient concentrations, diffusion constants, uptake rates and genetic oscillation cycle times $T_O$, see Figs. S4-S6 for the patterning rules under varying growth conditions). When gene oscillators in cells stop oscillating in the very early developmental stage of populations (see the evolution of the growth rate of frontier cells $\lambda_{front}$ in Fig. 6c1, where the radial distribution of $\lambda_C$ in the population is in the low mode as in Fig. 6d), sector-like gene patterns are produced. The patterning principle of sectors is mainly the local genotype fixation arising from the self-proliferation of cells, which further leads to the establishment of cell lineages with the same cell fates at a large space scale (Movie S6). When oscillators can persist for a long enough time and in the meantime the graded position information can be provided, circular fringes are produced (Fig. 6c4, Movie S7), which are in line with the previous experimental observation [21]. The underlying patterning mechanism is similar to the pattern formation of cell cycles in Fig. 2b. The synchronization between gene expressions in pioneering cells of the population is guaranteed by the radially symmetrical position information that is produced by the radially graded nutrient distribution (Fig. 2b3). When oscillation persists for a specific time and then ceases, a synthesis of sectors and circular fringes will be produced (Figs. 6c2,3).

We provide further support for the mode transition of gene patterns for the same growth condition through modulating the parameter $\lambda_{stop}$ individually. With the decrease in $\lambda_{stop}$, the mode of gene patterns transitions from sectors to circular stripes (Fig. S7).

Our results demonstrate that the stripe pattern formation requires the stable establishment of the graded radial distribution of $\lambda_C$ that crosses $\lambda_{stop}$ (i.e. the hybrid distribution mode in Fig. 6d). Besides, the genetic oscillation cycle time $T_O$ should also be long enough to obtain gene patterns with recognizable wave lengths, as there is a positive correlation between the wave length and $T_O$ (Fig. S8). In theory, $T_O$ should be larger than the patterning period of cell cycles (~ 1.1 h shown in Fig. 4d).

To have a direct knowledge about how cell cycle patterning affects genetic oscillation dynamics, we present in Fig. 8 the simultaneous oscillations of cell cycles, intracellular araC and lacI in cells located in different parts of the colony under different growth conditions but the same gene expression dynamics. The result suggests that there is a positive correlation between the spatial patterning of cell cycles and that of genetic kinetics. The larger the wave length of the cell size stripe is, the larger the wave length of the genetic stripe.

## 2.4 Gene pattern with bi-stability

The precise positioning of gene expression domains in cell populations is a big challenge for the current understanding of developmental pattern formation, since the continuously graded morphogen gradients generated by diffusion are believed to be too messy to drive the tissue patterning in organisms [36,37]. The typical case is the zebra stripe, where pigment cells have only two fates (white or black) and are strictly segregated in space. A single gene-oscillatory system struggles to produce such periodic patterns with discontinuous (discrete) pigment domains (see Fig. 6). An early guess for this problem is that different pigment genes are activated by a graded morphogen distribution that provides different position information in a concentration-dependent manner with corresponding thresholds, then cells adopt different fates at the signal levels which are below and above the thresholds (the French Flag model [38]). However, it is still a puzzle for this model why one gene cannot be activated on the higher morphogen (transcriptional factor) concentration if it can be activated at a lower level of concentration. It is also hard to explain the periodicity of discrete domains like in zebra stripes, with the French flag model. Moreover, although many studies on early embryonic development have suggested that geometrical constraints, genetic regulatory feedback, apoptosis, and cell-cell and cell-environment interactions are responsible for embryonic self-organization [39-42], the self-assembly process of cell-fate-specific domains in developing cell populations has still remained poorly understood in basic principles [43].

The inhibitor-inhibitor (II) mechanism has been found to be able to exclusively lock cells' fates into one of two fixed states (activation or silence) in an irreversible way to realize cell-fate decisions [44,45]. More recently, in an in-vivo study the necessity of a mutually inhibitory gene network consisting of Wnt and Sox9 was observed in the precise digit patterning of mouse [46]. We next investigate the formation mechanism of gene patterns with zebra stripes in expanding populations of discrete cells, based on the combined AAI-II gene-regulatory network shown in Fig. 9a. In our simulations, the II subsystem is not started until a cell receives a starting signal $\lambda_C < \lambda_{II,0}$. Our methodology for the current intracellular gene expression dynamics and parameters used are given in Fig. 10.

We observe that with the current genetic regulatory mechanism, periodic stripe patterns with only two states (C or D) are produced without having to shut down the gene-oscillatory subsystem consisting of the AAI network (Fig. 9b, Movies S8 and S9). How such gene patterns with discrete cell-state-specific domains can be established, according to the messy spatial distribution of state transition signals (A and B) and also the continuously graded position information provided by the nutrient distribution within the population? In our study, we found that in the development of cell populations, the operation of AAI subsystem in cells can stably provide the alternative signal level (high A or high B) for the cells around a given position $\lambda_C = \lambda_{II,0}$ (Fig. 9c, Movie S10), despite the spatial disorder of signals A and B within the whole cell population (Movie S11). Since the II subsystem locks a cell into one specific fate in a wide range of relative input signal level (not based on the absolute values of signals), a little difference in the oscillation cycle phases of transition signals inside those cells would not have a big impact on the coordination among their fate decisions. In this case, according to the alternative fate switching signals moving like traveling waves, all the cells, who receive a high level of A or B signal around that radial place, then make the same fate decision and precisely self-organize into the specific C or D pigment region by means of the II regulatory system (Fig. 9c). This also suggests that the fate region width scales with the expansion speed of cell populations and also the oscillation cycle time of state transition signals.

The precision of region-specific cell fate positioning is influenced by the parameter $\lambda_{II,0}$ (Fig. 9e). This is because that higher $\lambda_{II,0}$ means the fate decision-making at the position closer to the colony front, where cells have a higher velocity for free motions (Fig. 9d). In this case, the coordination in the location of cells receiving the A or B signal at the place will be reduced and the degree of mixing between pigment C and D cells is subsequently enhanced, suggesting the importance of slow free motions of cells in precise biological patterning (in embryonic development, slow free motions of cells correspond, to a large degree, to cell adhesion [41]). This result may provide important implications for us to understand the embryonic development of multicellular organisms, such as the development of animal spines.

## 3. Summary and discussion

At present, the elementary self-organization mechanism of biological pattern formation is widely believed as the local self-activation and lateral long-range inhibition, which is derived from a continuum theory for only two interacting and diffusing chemicals (i.e. the Turing model). Although this mechanism has become a tenet in biology for several decades as it spontaneously drives pattern formation, it might be inaccurate for complex biological development which usually involves multi-physics and multi-scale processes.

In case of *E. coli* strains, we develop a first-principle computational model for the development of cell populations, which is able to incorporate individual cellular behaviors on different time and space scales including cell growth and division, cell movement and intracellular gene expressions (compared with the Turing model, this approach is more precise but needs a huge amount of computations). We show here that spontaneous biological patterning within the cell populations developed from one single cell is driven by the graded position information that is produced by the nutrient depletion, with activators and inhibitors not required to diffuse through cell populations (activators not required to be self-enhanced, either). According to the moving positional information produced by diffusion, cell cycles and periodic gene expressions can self-organize into stripe structures within cell populations, with some necessary conditions satisfied (the establishment of genetic stripes requires the stable graded radial distribution of $\lambda_C$ that crosses $\lambda_{stop}$). Besides, we also observe that under unlimited nutrients, the cell cycle distributions produce periodic transitions between different modes of cell cycle. However, this behavior is completely different in nutrient-deficient conditions where a scale-invariant dependence of cell cycle distributions emerge and can be described by a 4-parameter empirical function.

Another important developmental question is about the precise establishment of specific gene expression domains within a cell population. Although the early French Flag model proposed that this problem can be solved if different genes are activated according to corresponding thresholds, it might have problems in its self-consistency. Here we show that with the involvement of an II gene-regulatory network that is activated by a gene-oscillatory network, cells are able to interpret messy gradients of fate transition signals into region-specific cell fates. A key secret behind is that cells make the same fate decision according to the relative not the absolute values of signals. Our results also show that the speed of free motions of cells has a significant impact on the precise cell fate positioning. These findings may provide us deeper insights into the relevant natural biological pattern formation.

The next step of our work is to include more details of subcellular events and cellular signaling network, such as the gene regulatory mechanism at the molecular level and the cell-cell signaling mediated via molecules, or even electrical and mechanical signals. These details may improve our understanding of complex pattern formation greatly.

**Acknowledgements:** This work is supported by Zhejiang University, National Natural Science Foundation of China and China Postdoctoral Science Foundation (grant No. 2019M662027). We are grateful to comments and improvement suggestions from reviewers.

**Author contributions**: X.W. conceived this research, developed models, implemented theoretical calculations and data analysis, and wrote the paper. D.B. provided revision suggestions and contributed to the writing. All authors participated in discussion.

**Conflict of interest**: The authors declare no conflict of interest.

**Data and materials availability**: All calculation data are available from X.W. upon reasonable request. We state that we have mainly used the experimental dataset from a previous publication [23] for our model calibration.

Table 1: Parameters defined in the main text.

| Parameter | Description | Value/Range | Unit |
|---|---|---|---|
| $C$ | Nutrient concentration | $0 \sim C_0$ | $K_S$ |
| $C_0$ | Initial nutrient concentration in solid agar | $1 \sim 10$ | $K_S$ |
| $C_{front}$ | Nutrient concentration at colony front | - | $K_S$ |
| $C_{center}$ | Nutrient concentration at colony center | - | $K_S$ |
| $D_s$ | Nutrient diffusion constant | $6 \times 10^4$ [18] | $\mu m^2 \cdot h^{-1}$ |
| $K_S$ | Nutrient concentration at which $\lambda_C = 0.5\lambda_S$ | - | - |
| $k_c$ | Critical proportion | - | - |
| $l_0$ | Initiation length of one cell cycle | 0.8 | $\mu m$ |
| $l_C$ | Cell length | $0.8 \sim 1.6$ | $\mu m$ |
| $l_{div}$ | Cell division length | 1.6 | $\mu m$ |
| $L_\lambda$ | Wave length of cell size stripes | - | $\mu m$ |
| $v_n$ | Moving speed of nutrient distribution | - | $\mu m \cdot h^{-1}$ |
| $T_O$ | Genetic oscillation cycle time | - | h |
| $\alpha$ | Scale factor correlating nutrient uptake rate | $0.1 \sim 2$ | - |
| $\Delta t_G$ | Time step for cell growth | $10^{-2}$ | h |
| $\Delta t_E$ | Time step for gene expression | $10^{-4}$ | h |
| $\Delta t_M$ | Time step for cell movement | $10^{-3}$ | h |
| $\Delta t_N$ | Time step for nutrient diffusion | $10^{-4}$ | h |
| $\lambda_C$ | Growth rate | - | $h^{-1}$ |
| $\lambda_{front}$ | Growth rate at colony front | - | $h^{-1}$ |
| $\lambda_S$ | Ideal steady-state growth rate | 1.4 [23] | $h^{-1}$ |
| $\lambda_{stop}$ | Growth rate below which genetic oscillations stop | - | $h^{-1}$ |
| $\lambda_{II,0}$ | Growth rate below which II gene network works | - | $h^{-1}$ |
| $\varphi_C$ | Cell cycle phase | - | rad |

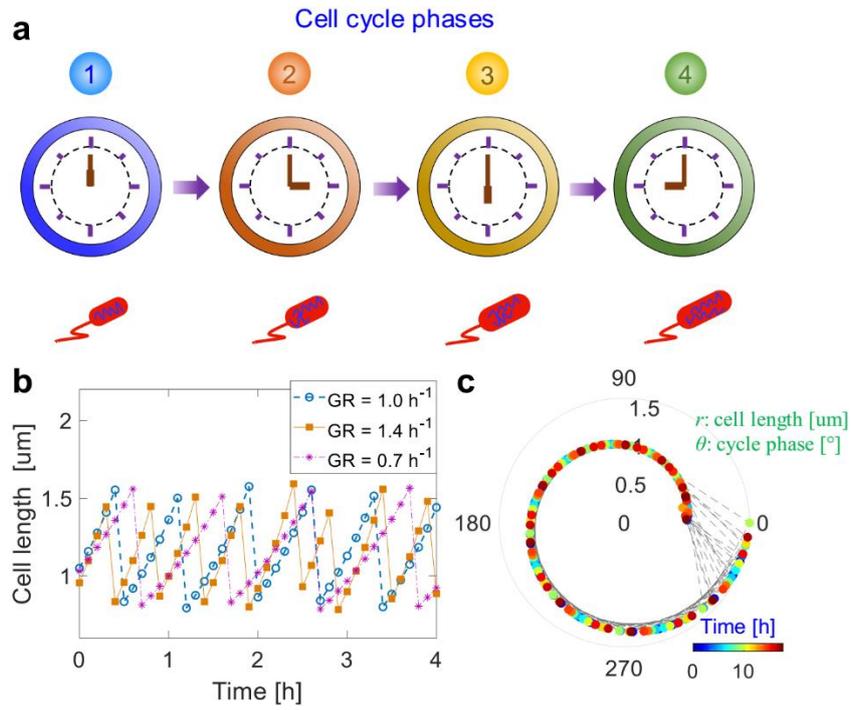

**Figure 1.** (**a**) Illustration of the cell growth and division cycle (in case of *E. coli*). (**b**) Variation of cell length with time under different growth rates (GRs, the standard growth rate of *E. coli* is about 1.0 h$^{-1}$ [27]). (**c**) Evolution of cell length in polar coordinates (GR = 0.7 h$^{-1}$). Variation of cell length in one cell cycle: $\partial l(t)/\partial t = \lambda_l l(t)$, $\lambda_l$ is the elongation rate, which is set to be equal to the growth rate here. $l_0 = 0.5 l_{div} = 0.8$ μm. Cell cycle time $T_{div}$ = ln2/GR.

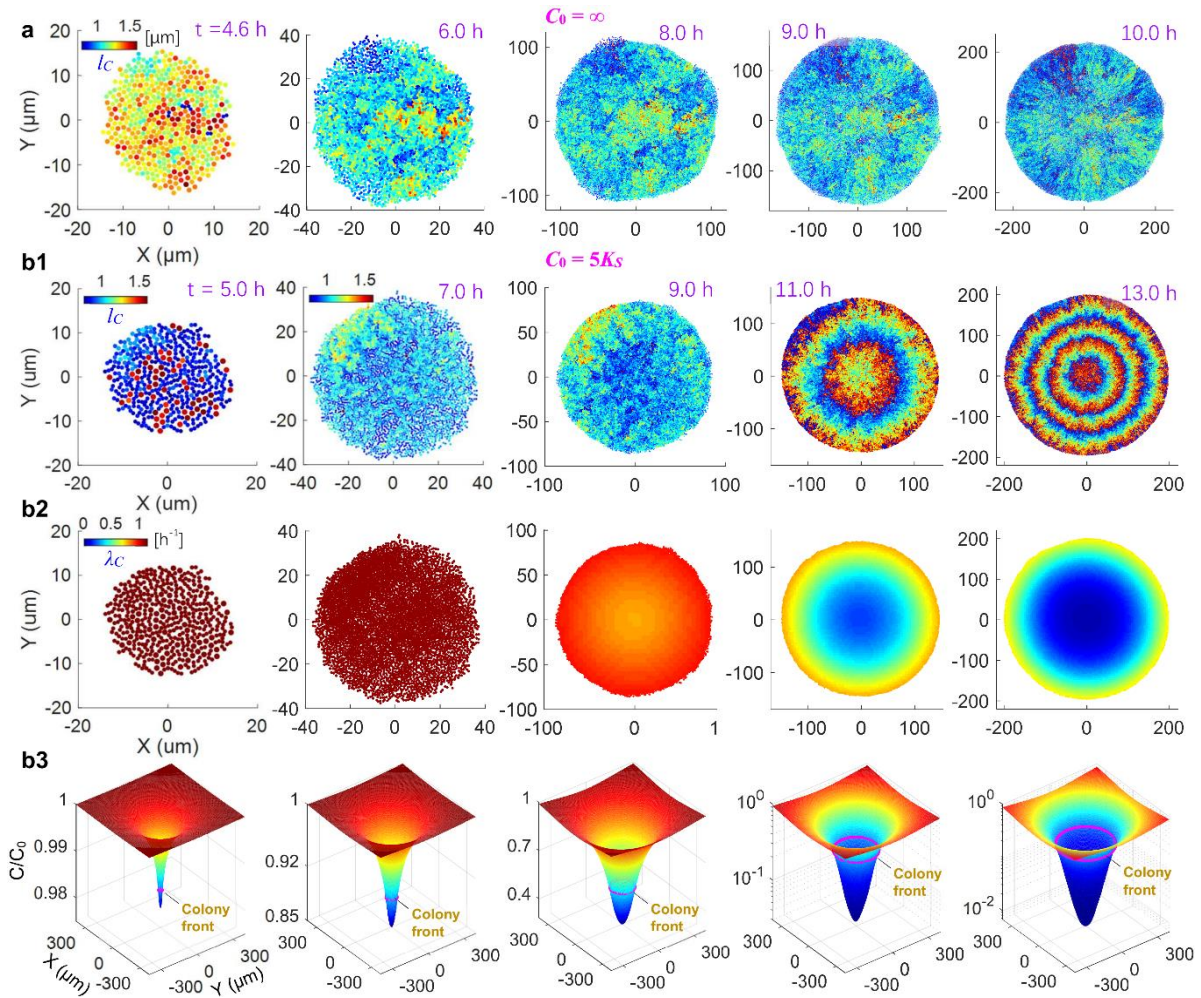

**Figure 2. Self-organization of cell cycles (illustrated with cell lengths) within the expanding cell population.** (**a**) Unlimited nutrients: the growth rate of all cells is constantly equal to $\lambda s$ = 1.4 h$^{-1}$ over time. (**b1-b3**) Limited nutrients: initial nutrient concentration in solid agar $C_0 = 5K_S$ ($K_S$ denotes the nutrient concentration at which the cell growth rate reduces to 0.5 $\lambda s$); (**b1**) Evolution of spatial arrangement of cell lengths $l_C$; (**b2**) Evolution of spatial distribution of cell growth rates $\lambda_C$; (**b3**) Evolution of spatial distribution of nutrient concentration. $\lambda s$ = 1.4 h$^{-1}$, nutrient diffusion constant $D_s$ = 6×10$^4$ μm$^2$/h [18], and the scale factor of nutrient uptake rate α = 1 [29].

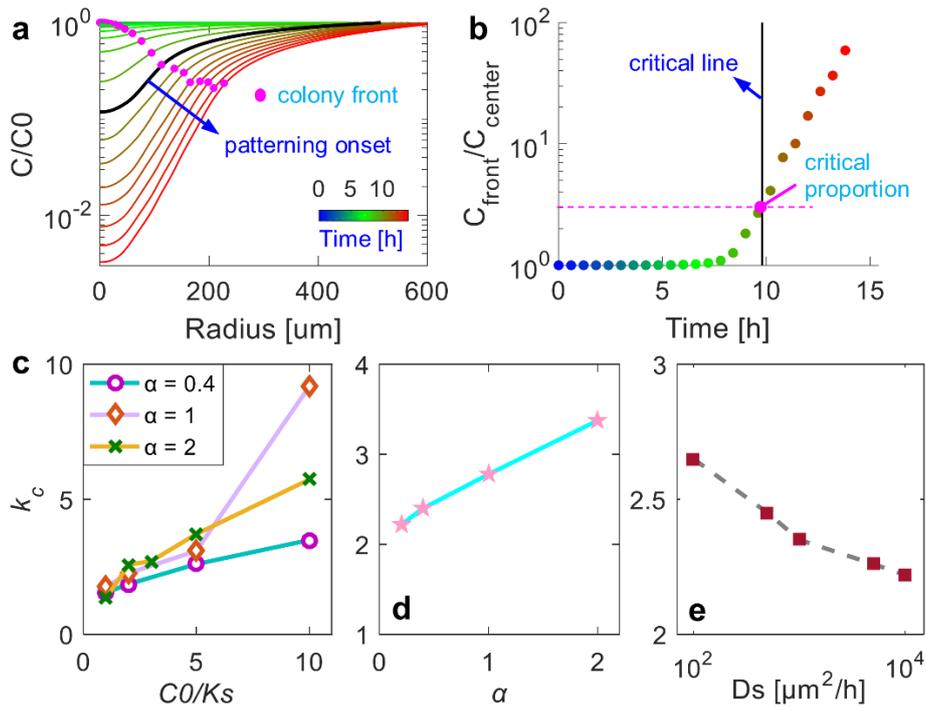

**Figure 3.** (**a**) Evolution of radial nutrient distribution ($D_s = 6\times10^4$ μm$^2$/h, $\alpha = 1$, $C_0/K_s = 5$). (**b**) Evolution of the proportion $C_{front}/C_{center}$. (**c**)-(**e**) Variation of critical proportion of nutrient gradient for patterning $k_c$ with influencing factors: (**c**) Effect of nutrient concentration ($D_s = 6\times10^4$ μm$^2$/h); (**d**) Effect of nutrient uptake rate (tuned via the scale factor $\alpha$, $D_s = 1\times10^4$ μm$^2$/h, $C_0/K_s = 5$); (**e**) Effect of nutrient diffusion coefficient ($\alpha = 1$, $C_0/K_s = 5$).

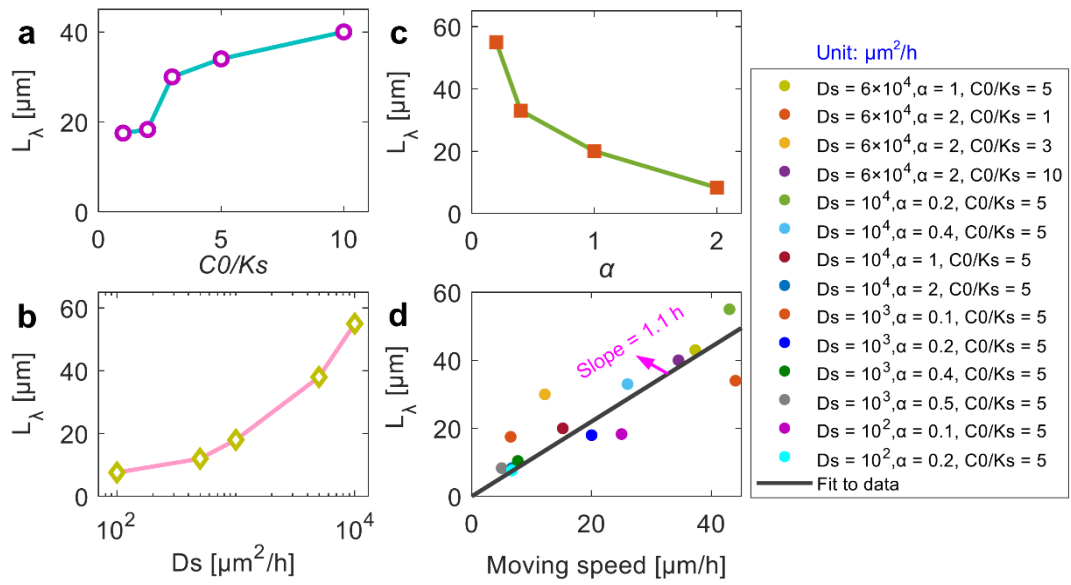

**Figure 4. Variation of the wave length of cell size stripes $L_\lambda$ with influencing factors.** (**a**) Effect of nutrient concentration ($D_s$ = 6×10$^4$ μm$^2$/h, $\alpha$ = 1); (**b**) Effect of nutrient diffusion coefficient ($\alpha$ = 1, $C_0/K_S$ = 5); (**c**) Effect of nutrient uptake rate (tuned via the scale factor $\alpha$, $D_s$ = 1×10$^4$ μm$^2$/h, $C_0/K_S$ = 5); (**d**) Relationship between the wave length of cell size stripes $L_\lambda$ and the nutrient moving speed.

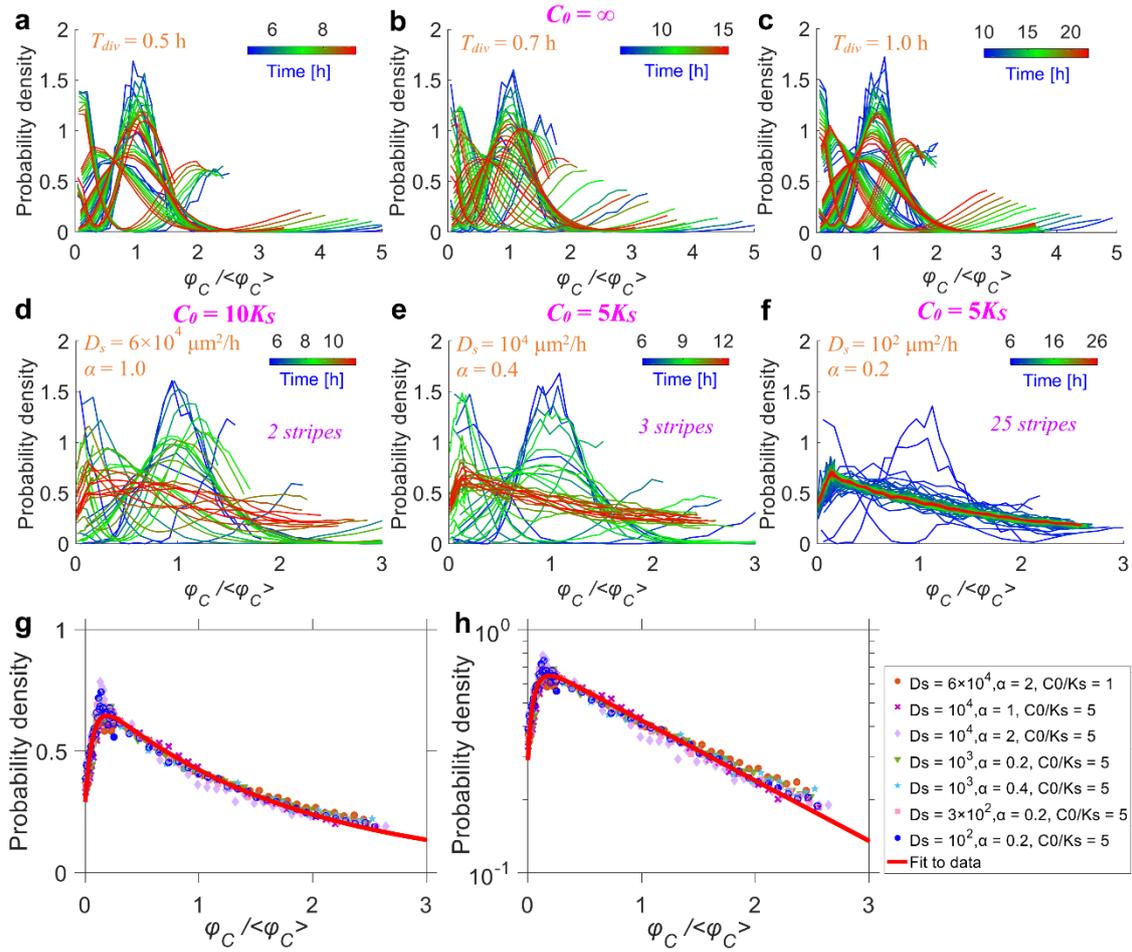

**Figure 5. Evolution of statistical distributions of cell cycle phases within the population.** (**a**)-(**c**) Unlimited nutrients: (**a**) Cell doubling time $T_{div}$ = 0.5 h; (**b**) $T_{div}$ = 0.7 h; (**c**) $T_{div}$ = 1.0 h. (**d**)-(**f**) Limited nutrients: (**d**) $C_0 = 10K_S$ ($D_s = 6\times10^4$ μm²/h, $\alpha$ = 1); (**e**) $C_0 = 5K_S$ ($D_s = 10^4$ μm²/h, $\alpha$ = 0.4); (**f**) $C_0 = 5K_S$ ($D_s = 10^2$ μm²/h, $\alpha$ = 0.2). (**g**),(**h**) The universal statistical distribution of cell cycle phases within cell populations: Plotted on (**g**) linear and (**h**) semi-logarithmic axes.

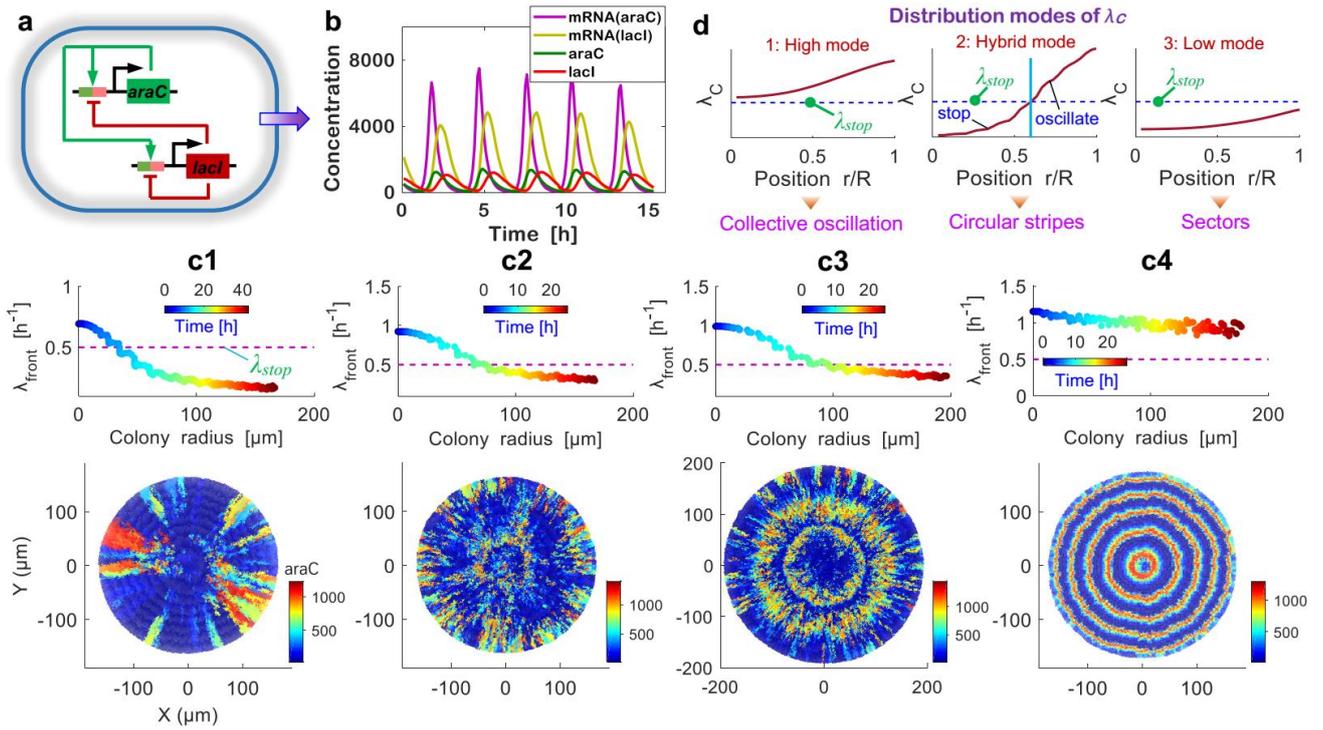

**Figure 6. Genetic pattern formation in the development of cell populations.** (**a**) Genetic oscillatory circuits with positive-negative feedbacks in single cells. (**b**) Evolution of mRNAs and proteins in cell (cell 1, the ancestor cell). (**c1**)-(**c4**) Genetic patterns: (**c1**) Sector-like patterns ($D_s$ = 6×10$^4$ μm$^2$/h, $\alpha$ = 2.0, $C_0/K_S$ = 1, $T_0$ = 2.2 h); (**c2**) Ring-Sector pattern (one stripe, $D_s$ = 6×10$^4$ μm$^2$/h, $\alpha$ = 2.0, $C_0/K_S$ = 2, $T_0$ = 1.1 h); (**c3**) Ring-sector pattern (two stripes, $D_s$ = 6×10$^4$ μm$^2$/h, $\alpha$ = 2.0, $C_0/K_S$ = 2.5, $T_0$ = 2.2 h); (**c4**) Ring pattern ($D_s$ = 10$^2$ μm$^2$/h, $\alpha$ = 0.2, $C_0/K_S$ = 5, $T_0$ = 4.4 h). (**d**) Radial distribution schemes of cell growth rates within the cell population: (**1**) High mode: the whole distribution of $\lambda_C$ is in high level that is above $\lambda_{stop}$; (**2**) Hybrid mode: the whole distribution of $\lambda_C$ is graded that crosses $\lambda_{stop}$; (**3**) Low mode: the whole distribution of $\lambda_C$ is in low level that is below $\lambda_{stop}$. Here $\lambda_{stop}$ = 0.5 h$^{-1}$.

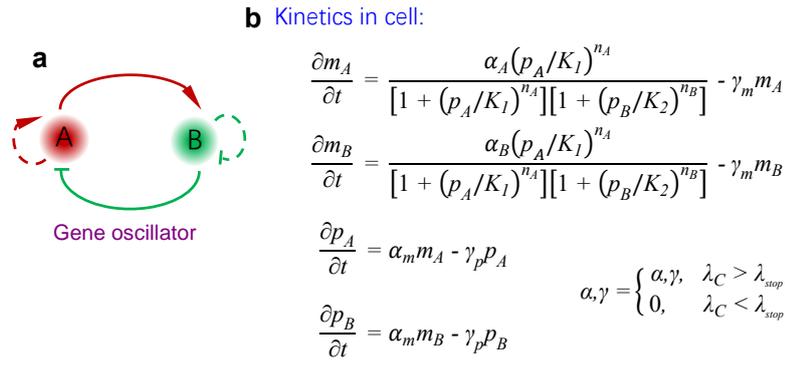

**a** Gene oscillator

**b** Kinetics in cell:

$$\frac{\partial m_A}{\partial t} = \frac{\alpha_A (p_A/K_1)^{n_A}}{[1+(p_A/K_1)^{n_A}][1+(p_B/K_2)^{n_B}]} - \gamma_m m_A$$

$$\frac{\partial m_B}{\partial t} = \frac{\alpha_B (p_A/K_1)^{n_A}}{[1+(p_A/K_1)^{n_A}][1+(p_B/K_2)^{n_B}]} - \gamma_m m_B$$

$$\frac{\partial p_A}{\partial t} = \alpha_m m_A - \gamma_p p_A$$

$$\frac{\partial p_B}{\partial t} = \alpha_m m_B - \gamma_p p_B$$

$$\alpha,\gamma = \begin{cases} \alpha,\gamma, & \lambda_C > \lambda_{stop} \\ 0, & \lambda_C < \lambda_{stop} \end{cases}$$

**c**

Table 2: Parameters in genetic dynamics.

| Parameter | Description | Value/Range | Unit |
|---|---|---|---|
| $K_1$ | Hill constant | 1000 | - |
| $K_2$ | Hill constant | 30 | - |
| $n_A$ | Hill exponent | 2 [34] | - |
| $n_B$ | Hill exponent | 4 [34] | - |
| $\alpha_A$ | Synthesis rate | $5\times10^4$ | $h^{-1}$ |
| $\alpha_B$ | Synthesis rate | $5\times10^4$ | $h^{-1}$ |
| $\alpha_m$ | Synthesis rate | 1 | $h^{-1}$ |
| $\gamma_m$ | Degeneration rate | 4 | $h^{-1}$ |
| $\gamma_p$ | Degeneration rate | 3 | $h^{-1}$ |

**Figure 7.** (**a**) Genetic topology for the activator-inhibitor network; (**b**) Mathematical description of the intracellular gene expression; (**c**) Parameters used for genetic kinetics.

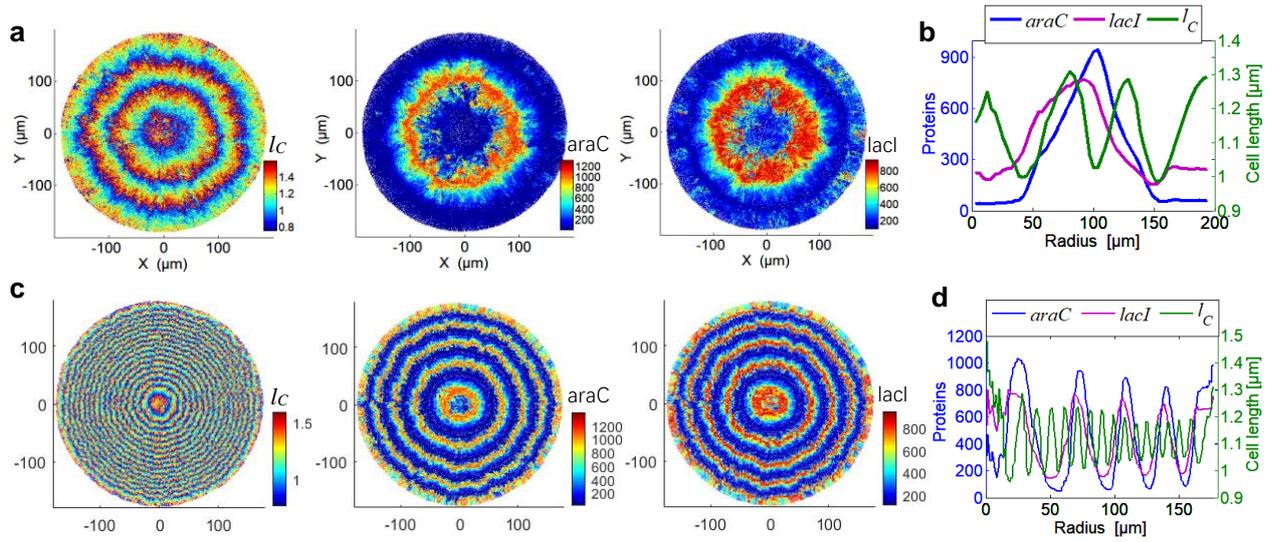

**Figure 8. Pattern formation of cell cycles and gene expressions in the development of cell populations** (**a**),(**b**) $D_s = 6\times10^4$ μm$^2$/h, $\alpha = 1.0$, $C_0/K_S = 5$, To = 3.1 h: (**a**) Spatial distributions of cell lengths, intracellular araC and lacI; (**b**) Statistical radial distribution of cell lengths, araC and lacI. (**c**),(**d**) $D_s = 3\times10^2$ μm$^2$/h, $\alpha = 0.2$, $C_0/K_S = 5$, To = 3.1 h. (**c**) Spatial distributions of cell lengths, intracellular araC and lacI; (**d**) Statistical radial distribution of cell lengths, araC and lacI.

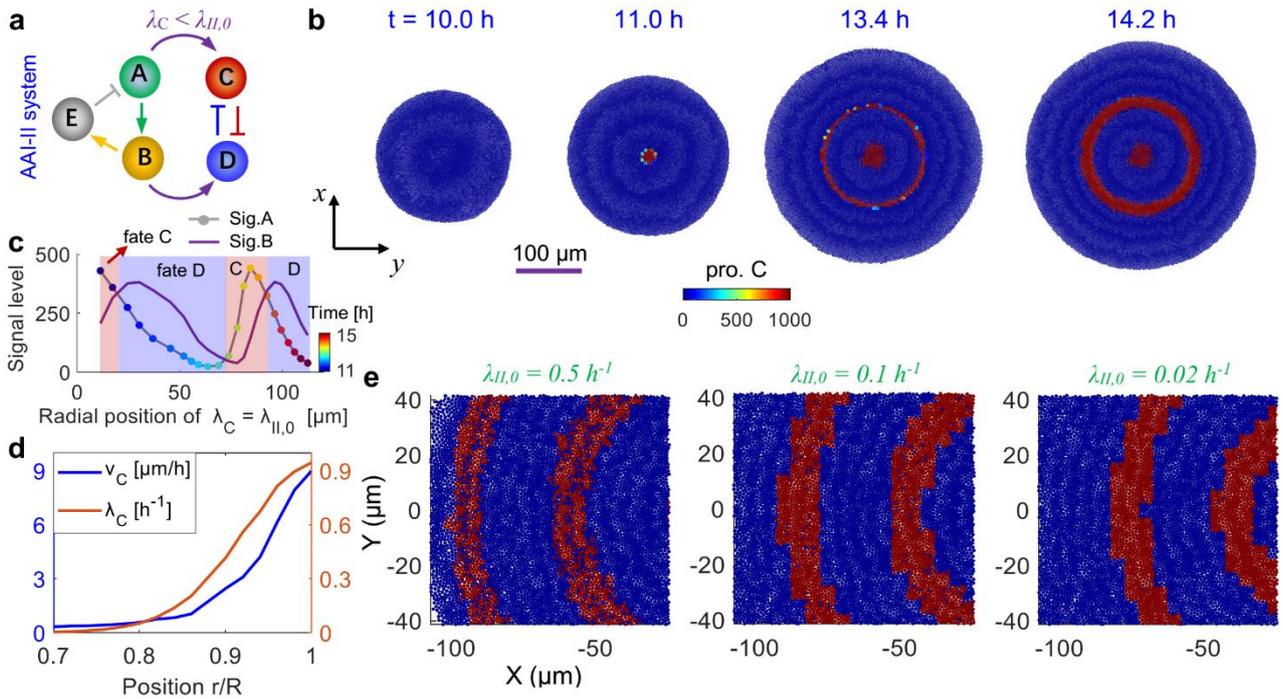

**Figure 9. Patterning of zebra stripes in the development of cell populations.** (**a**) The AAI-II gene network system in cells. (**b**), (**c**) Formation principle of the gene pattern with bi-stability ($D_s = 10^4$ μm$^2$/h, $\alpha = 0.5$, $C_0/K_S = 5$, $T_0 = 3.1$ h, $\lambda_{II,0} = 0.1$ h$^{-1}$): (**b**) Patterning process (shown with the spatial distribution of the protein C); (**c**) Evolution of alternative signals A and B at the position that cells start to make fate decisions: fate transition information moves like a traveling wave. (**d**), (**e**) Influence of the parameter $\lambda_{II,0}$ on the precision of region-specific fate positioning ($D_s = 10^2$ μm$^2$/h, $\alpha = 0.2$, $C_0/K_S = 5$, $T_0 = 4.4$ h): (**d**) Radial distribution of cells' motion speed $v_C$ and growth rate $\lambda_C$ (measured at the colony radius $R \approx 125$ μm); (**e**) Bi-stable gene patterns under different $\lambda_{II,0}$.

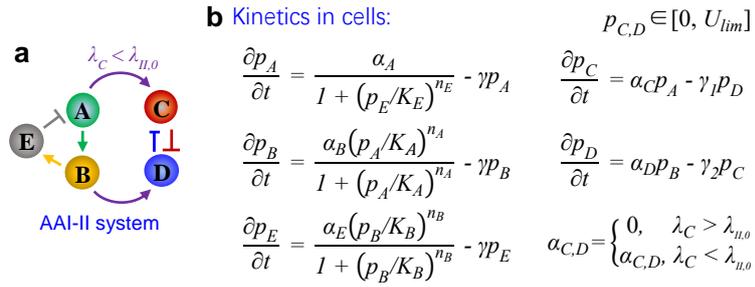

**b** Kinetics in cells:

$$\frac{\partial p_A}{\partial t} = \frac{\alpha_A}{1 + (p_E/K_E)^{n_E}} - \gamma p_A$$

$$\frac{\partial p_B}{\partial t} = \frac{\alpha_B (p_A/K_A)^{n_A}}{1 + (p_A/K_A)^{n_A}} - \gamma p_B$$

$$\frac{\partial p_E}{\partial t} = \frac{\alpha_E (p_B/K_B)^{n_B}}{1 + (p_B/K_B)^{n_B}} - \gamma p_E$$

$$p_{C,D} \in [0, U_{lim}]$$

$$\frac{\partial p_C}{\partial t} = \alpha_C p_A - \gamma_1 p_D$$

$$\frac{\partial p_D}{\partial t} = \alpha_D p_B - \gamma_2 p_C$$

$$\alpha_{C,D} = \begin{cases} 0, & \lambda_C > \lambda_{II,0} \\ \alpha_{C,D}, & \lambda_C < \lambda_{II,0} \end{cases}$$

**a** AAI-II system

**c**
Table 3: Parameters in genetic dynamics.

| Parameter | Description | Value/Range | Unit |
|---|---|---|---|
| $K_A$ | Hill constant | 2000 | - |
| $K_B$ | Hill constant | 2000 | - |
| $K_E$ | Hill constant | 10 | - |
| $n_A$ | Hill exponent | 2 | - |
| $n_B$ | Hill exponent | 2 | - |
| $n_E$ | Hill exponent | 4 | - |
| $U_{lim}$ | Upper limit value | 1500 | - |
| $\alpha_A$ | Synthesis rate | $5 \times 10^4$ | $h^{-1}$ |
| $\alpha_B$ | Synthesis rate | $5 \times 10^4$ | $h^{-1}$ |
| $\alpha_E$ | Synthesis rate | $1 \times 10^4$ | $h^{-1}$ |
| $\alpha_C$ | Synthesis rate | 100 | $h^{-1}$ |
| $\alpha_D$ | Synthesis rate | 100 | $h^{-1}$ |
| $\gamma$ | Degeneration rate | 4 | $h^{-1}$ |
| $\gamma_1$ | Degeneration rate | 100 | $h^{-1}$ |
| $\gamma_2$ | Degeneration rate | 100 | $h^{-1}$ |

**Figure 10.** (**a**) Genetic topology for the AAI-II network; (**b**) Mathematical description of the intracellular gene expression; (**c**) Parameters used for genetic kinetics.

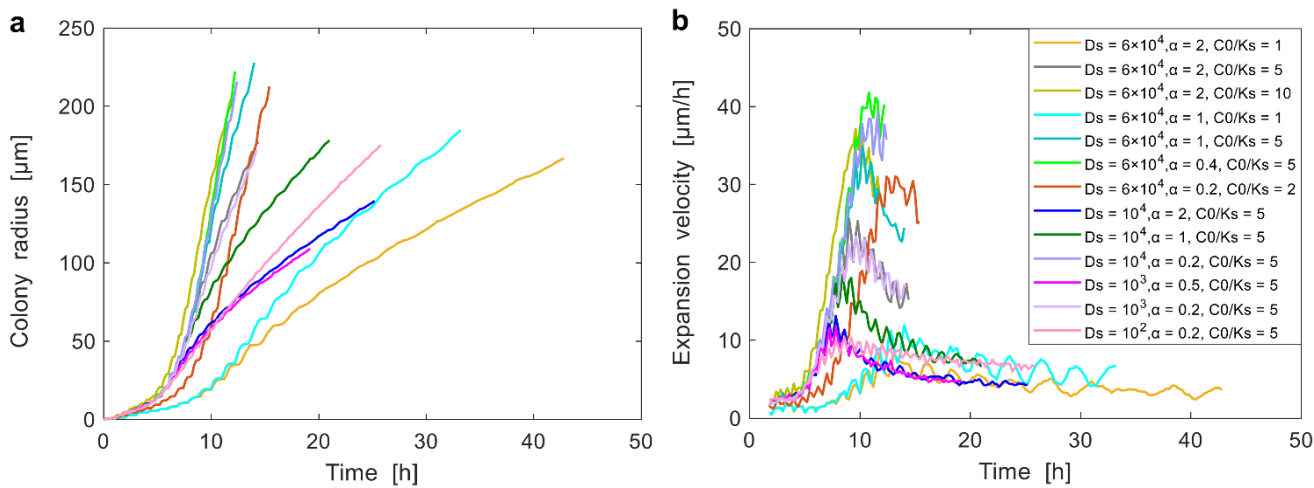

**Figure S1.** Evolution of colony radius (**a**) and expansion speed (**b**) under various growth conditions. Under limited nutrients, the range expansion of cell populations generally first undergoes an acceleration phase and then a deceleration phase. $\lambda_S$ = 1.4 h$^{-1}$.

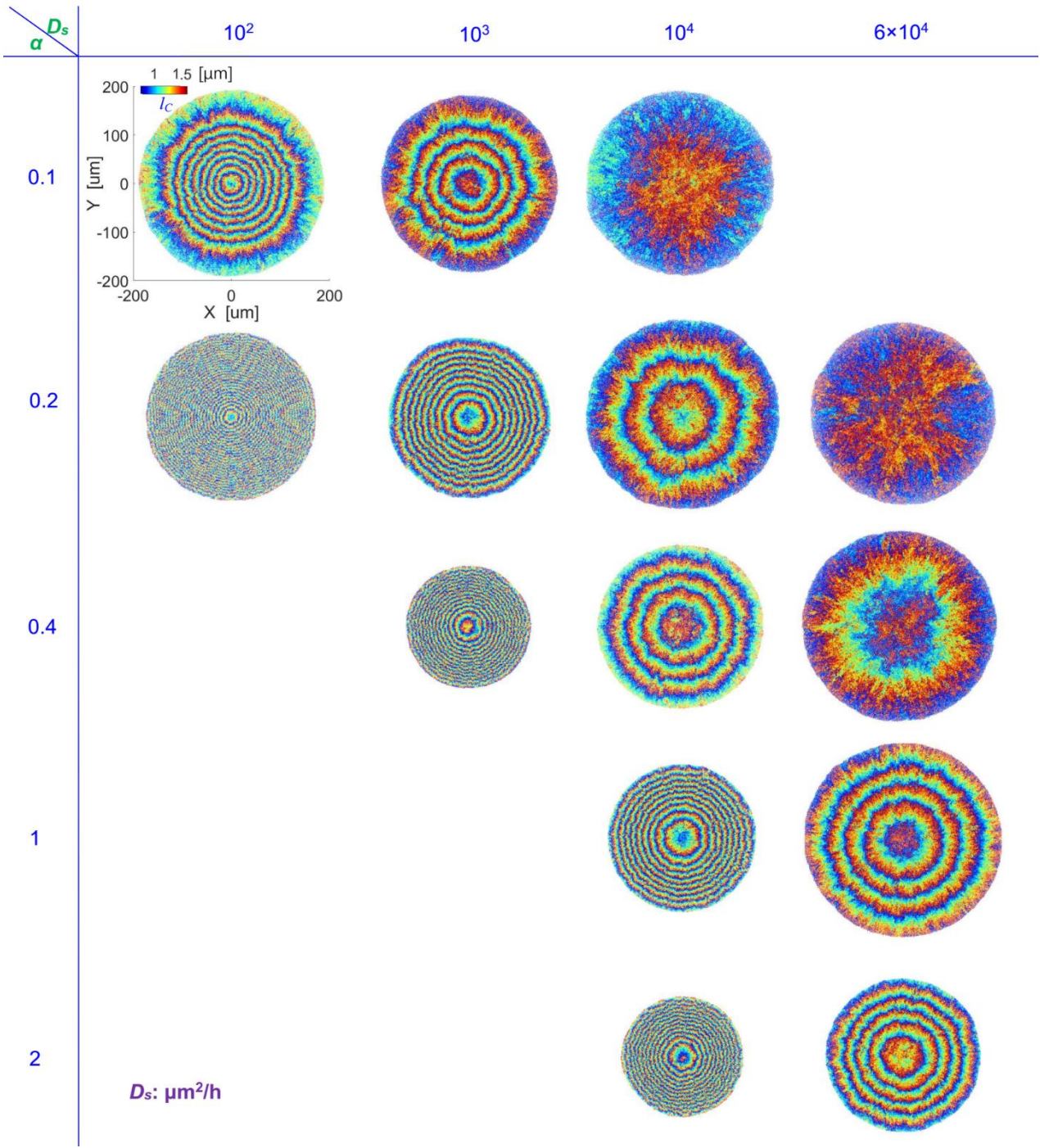

Figure S2. Pattern formation of cell cycles in developing cell populations for varying nutrient uptake rates (tuned via the scale factor $\alpha$) and diffusion constants $D_s$. $C_0/K_S = 5$.

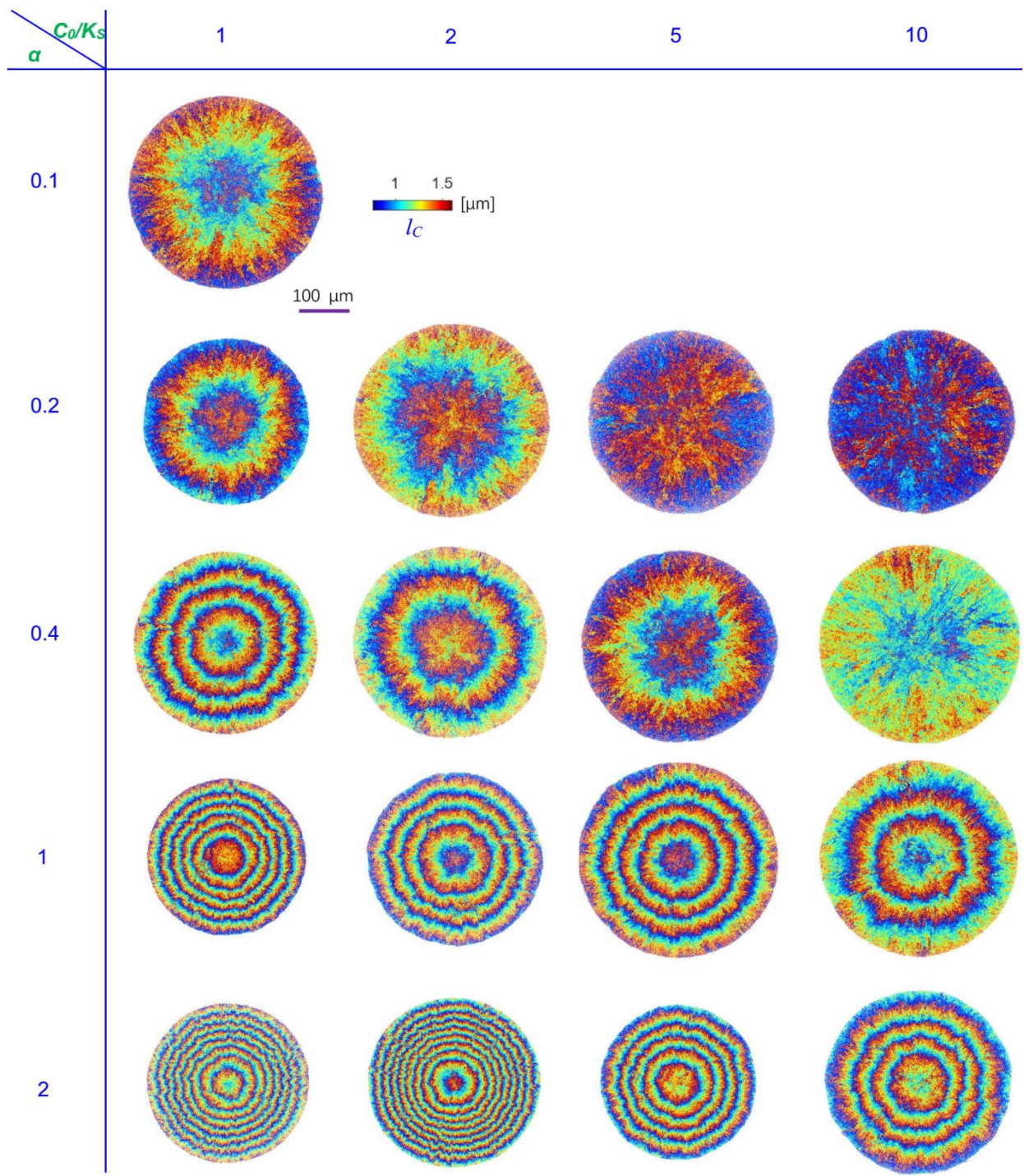

**Figure S3. Pattern formation of cell cycles in developing cell populations for varying nutrient uptake rates (tuned via the scale factor $\alpha$) and nutrient concentrations $C_0$.** $D_s = 6 \times 10^4$ μm²/h.

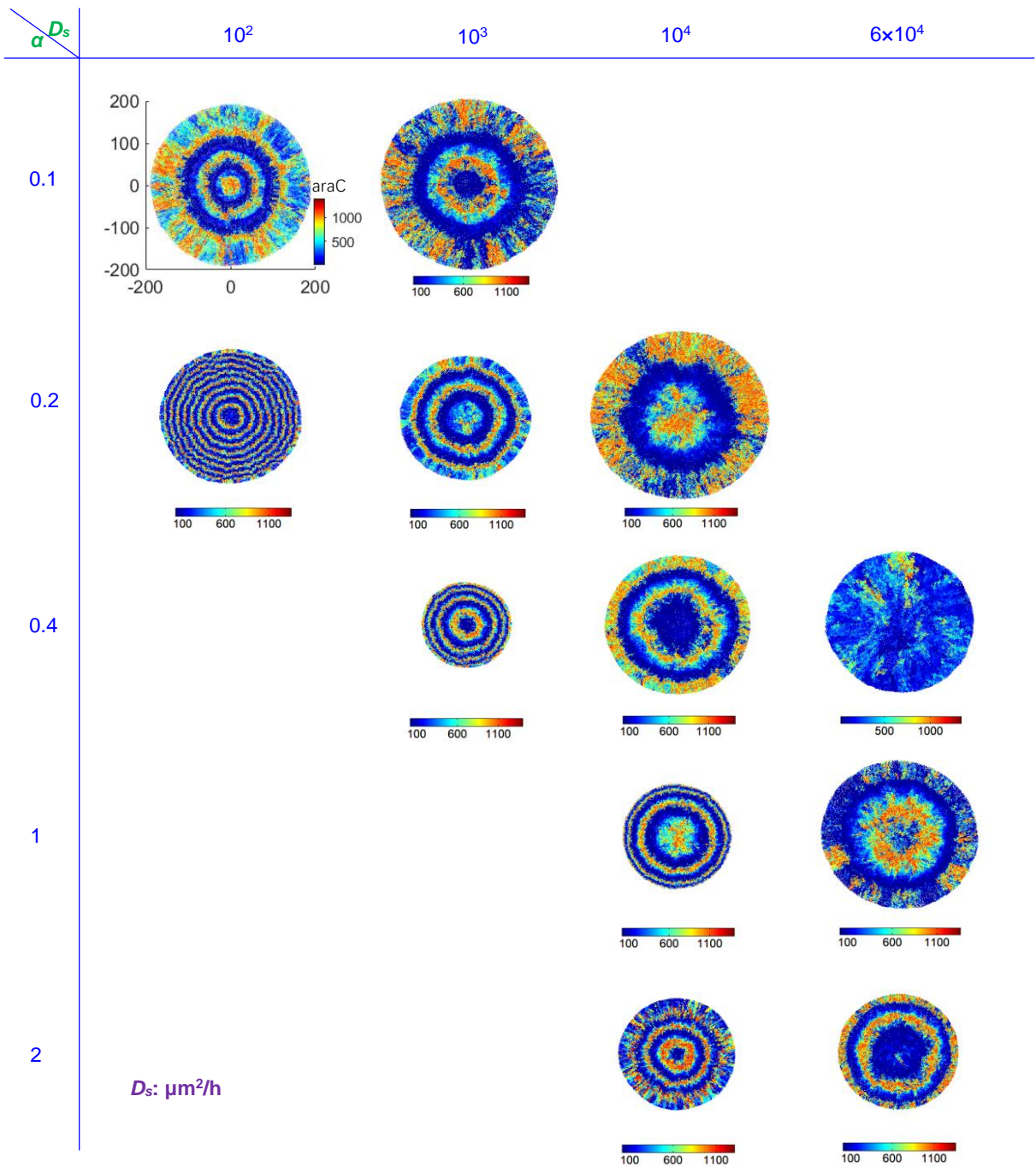

Figure S4. Genetic pattern formation in developing cell populations for varying nutrient uptake rates (tuned via the scale factor $\alpha$) and diffusion constants $D_s$. $C_0/K_S$ = 5 and $T_O$ = 2.2 h.

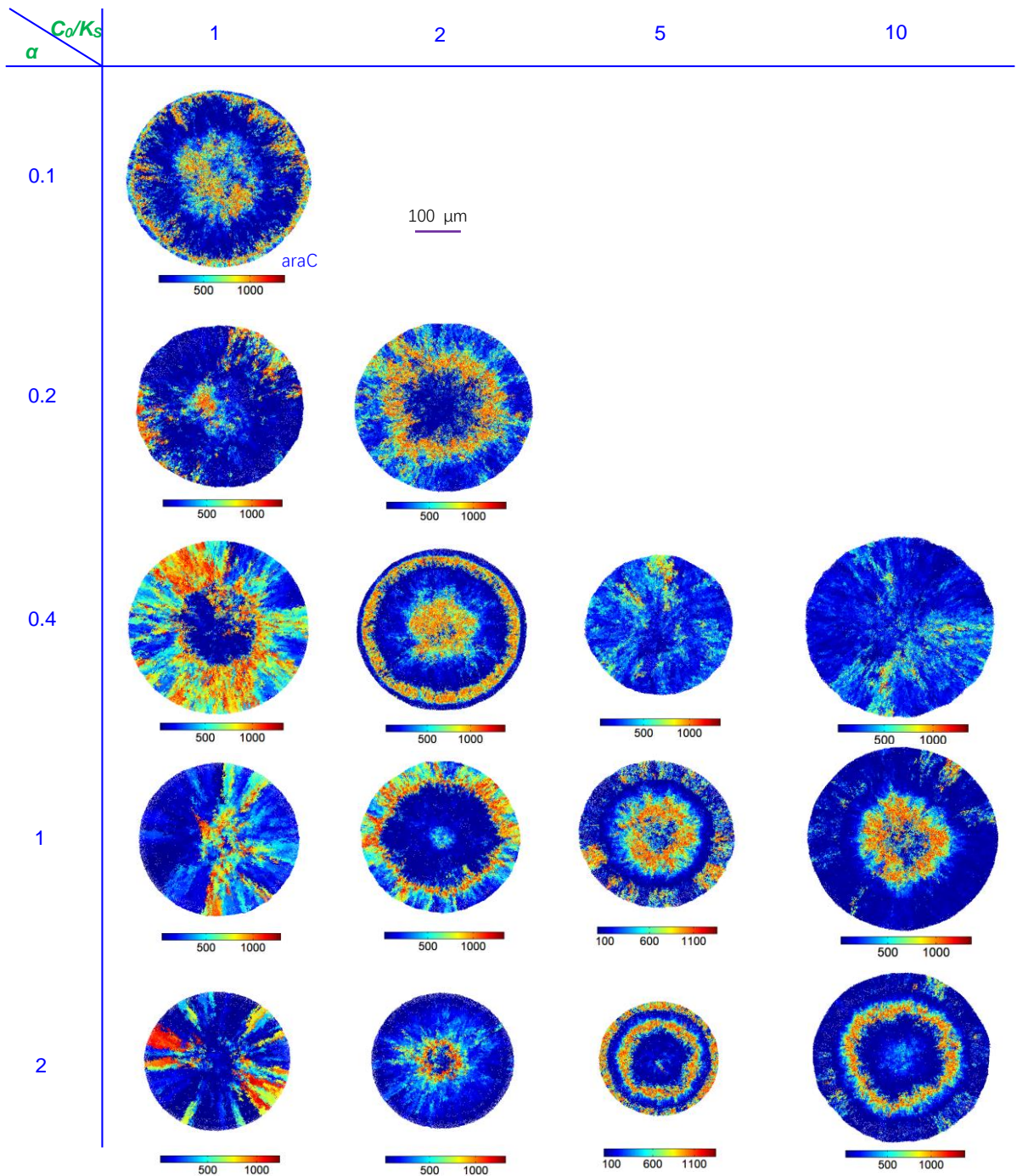

Figure S5. Genetic Pattern formation in developing cell populations for varying nutrient uptake rates (tuned via the scale factor $\alpha$) and nutrient concentrations $C_0$. $D_s = 6 \times 10^4$ µm$^2$/h and $T_o = 2.2$ h.

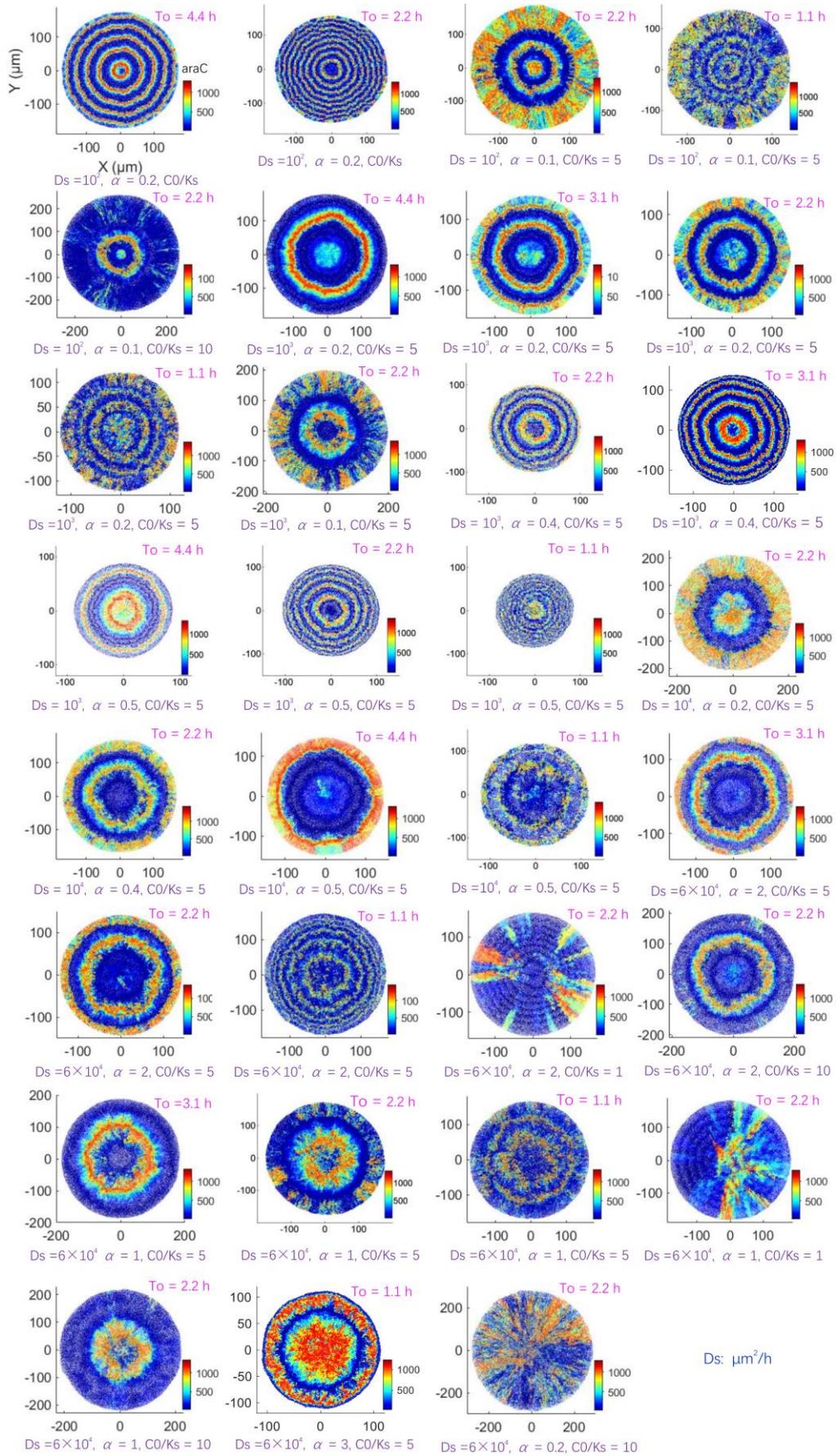

Figure S6. Genetic pattern formation in the development of cell populations under various growth conditions.

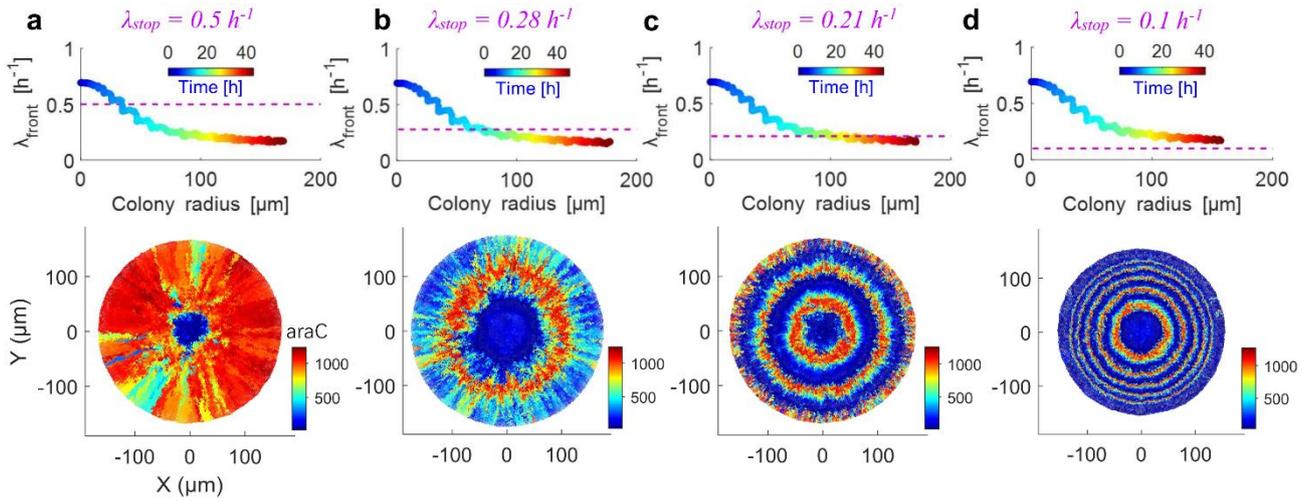

**Figure S7.** Genetic pattern formation in the development of cell populations under (**a**) $\lambda_{stop}$ = 0.5 h$^{-1}$, (**b**) $\lambda_{stop}$ = 0.28 h$^{-1}$, (**c**) $\lambda_{stop}$ = 0.21 h$^{-1}$ and (**d**) $\lambda_{stop}$ = 0.1 h$^{-1}$. $D_s$ = 6×10$^4$ µm$^2$/h, $\alpha$ = 2.0, $C_0/K_S$ = 1 and $T_o$ = 4.4 h.

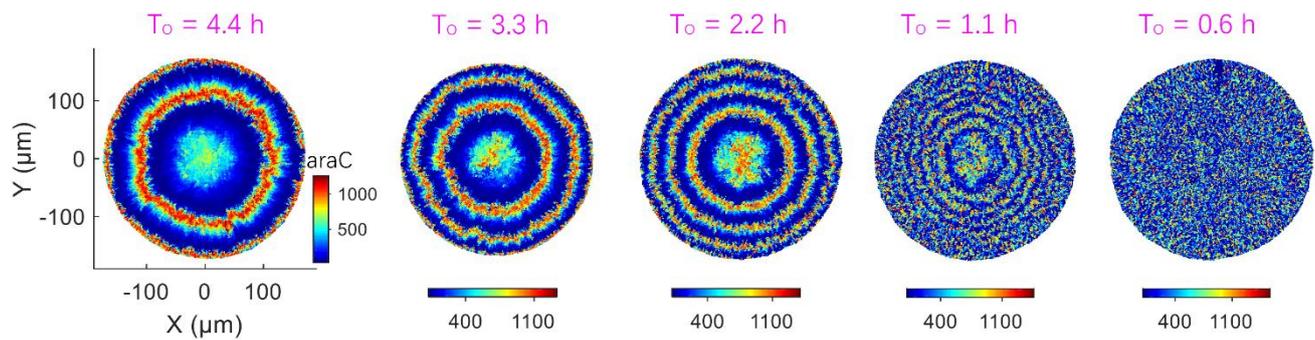

**Figure S8.** Genetic pattern formation under several genetic oscillation cycle times $T_o$ (illustrated with the spatial distribution of the intracellular protein araC). $D_s = 10^4$ μm$^2$/h, $\alpha = 1.0$, $C_0/K_S = 5$.

## Movie legends

**Movie S1: Pattern formation of cell cycles (shown with cell lengths) in the cell population developed from one single cell.** The built-in figure shows the evolution of radial distribution of cell growth rates in the population. When a sufficiently graded position information is provided, cell cycles start patterning. $D_s = 6\times10^4$ µm²/h, $\alpha = 1.0$ and $C_0/K_S = 5$.

**Movie S2: Evolution of cell growth rates in the cell population developed from one single cell.** The built-in figure shows the evolution of the expansion speed of the population and the moving speed of position information ($\lambda_C = 0.5$ h⁻¹). As the range expansion of the cell population approaches a steady state, the moving speed of position information will be close to the expansion speed. $D_s = 6\times10^4$ µm²/h, $\alpha = 1.0$ and $C_0/K_S = 5$.

**Movie S3: Evolution of the spatial distribution of nutrient concentration.** $D_s = 6\times10^4$ µm²/h, $\alpha = 1.0$ and $C_0/K_S = 5$.

**Movie S4: Evolution of the statistical distribution of cell cycle phases within the whole population: steady-state growth condition.** $\lambda_C \equiv 0.7$ h⁻¹.

**Movie S5: Evolution of the statistical distribution of cell cycle phases within the whole population: limited growth condition.** $D_s = 10^2$ µm²/h, $\alpha = 0.2$ and $C_0/K_S = 5$.

**Movie S6: Formation of the sector-like gene pattern.** The built-in figure shows the evolution of radial distribution of cell growth rates in the population $\lambda_c(r)$ and also the evolution of the growth rate of frontier cells $\lambda_{front}$. $\lambda_c(r)$ is below $\lambda_{stop}$ (in the low mode). $D_s = 6\times10^4$ µm²/h, $\alpha = 2.0$, $C_0/K_S = 1$ and $T_0 = 2.2$ h.

**Movie S7: Formation of the circular stripe gene pattern.** The built-in figure shows the evolution of radial distribution of cell growth rates in the population $\lambda_c(r)$ and also the evolution of the growth rate of frontier cells $\lambda_{front}$. $\lambda_c(r)$ is across $\lambda_{stop}$ (in the hybrid mode). $D_s = 10^2$ µm²/h, $\alpha = 0.2$, $C_0/K_S = 5$ and $T_0 = 4.4$ h.

**Movie S8: Establishment of the cell population with zebra stripes.** The built-in figure shows the evolution of fate transition signals A and B. According to the moving position information ($\lambda_{II,0} = 0.1$ h⁻¹), cells around that place will start the inhibitor-inhibitor (II) gene regulatory system to prepare for fate decision-making. Then according to levels of input signals A and B, a cell will decide to lock into the fate C or D. $D_s = 10^4$ µm²/h, $\alpha = 0.5$, $C_0/K_S = 5$ and $T_0 = 3.1$ h.

**Movie S9: Establishment of the cell population with zebra stripes.** Patterning principle is identical to that in movie 8. $D_s = 10^2$ µm²/h, $\alpha = 0.2$, $C_0/K_S = 5$, $T_0 = 4.4$ h and $\lambda_{II,0} = 0.1$ h⁻¹.

**Movie S10: Quasi-synchronous oscillation of fate transition signals in cells around the decision-starting place.** The built-in figure shows the evolution of fate transition signals A and B. Around the place where the position information is provided, cells can make precise fate decision according to the relative signal level rather than the signal value. In this case, even though there is some difference in oscillation cycle phases of transition signals, cells around the place are still able to keep good coordination in making the same fate decision. $D_s = 10^4$ µm²/h, $\alpha = 0.5$, $C_0/K_S = 5$ and $T_0 = 3.1$ h.

**Movie S11: Evolution of the intracellular signal A within the whole population.** The built-in figure shows the evolution of the signal A at the position where $\lambda_C = 0.1$ h⁻¹. The spatial distribution of signal A in the population is messy. $D_s = 10^4$ µm²/h, $\alpha = 0.5$, $C_0/K_S = 5$ and $T_0 = 3.1$ h.